\documentclass[journal]{IEEEtran}

\ifCLASSINFOpdf
\else
   \usepackage[dvips]{graphicx}
\fi
\usepackage{url}

\usepackage{graphicx}
\usepackage{multirow}
\usepackage{booktabs}
\usepackage{extra_styles}

\begin{document}

\title{Bridging the Gap between Audio and Text using Parallel-attention for User-defined Keyword Spotting}

\author{Youkyum Kim, Jaemin Jung, Jihwan Park, Byeong-Yeol Kim, and Joon Son Chung

\thanks{The first two authors contributed equally to this work.}

\thanks{Youkyum Kim, Jaemin Jung, and Joon Son Chung are with School of Electric Engineering, Korea Advanced Institute of Science and Technology, Daejeon 34141, Republic of Korea (e-mail:dbrua1998@kaist.ac.kr; jjm5811@kaist.ac.kr; joonson@kaist.ac.kr)

Jihwan Park and Byeong-Yeol Kim are with 42dot Inc., Seoul 06620, Republic of Korea (e-mail: jihwan.park@42dot.ai; byeongyeol.kim@42dot.ai)
}}

\maketitle

\begin{abstract}

    This paper proposes a novel user-defined keyword spotting framework that accurately detects audio keywords based on text enrollment. 
    Since audio data possesses additional acoustic information compared to text, there are discrepancies between these two modalities.
    To address this challenge, we present ParallelKWS, which utilises self- and cross-attention in a parallel architecture to effectively capture information both within and across the two modalities.
    We further propose a phoneme duration-based alignment loss that enforces the sequential correspondence between audio and text features. Extensive experimental results demonstrate that our proposed method achieves state-of-the-art performance on several benchmark datasets in both seen and unseen domains, without incorporating extra data beyond the dataset used in previous studies.

     
\end{abstract}

\IEEEpeerreviewmaketitle

\section{Introduction}

\IEEEPARstart{K}{eyword} spotting (KWS) plays a crucial role as an entry point for initiating voice-activated services on smart devices, which have recently been in growing demand. Earlier KWS systems~\cite{chen2014small, hou2019region, tang2018deep, berg2021keyword, huh2021metric} based on deep learning primarily focused on detecting only pre-defined keywords. With the rapid advancement of artificial intelligence services and the need for enhanced user experience, there has been a shift towards user-defined keyword spotting (UDKWS) systems. These systems allow users to set their own keywords, broadening the scope and applicability of KWS.

Previous works~\cite{chen2015query, huang2021query, mazumder21_interspeech, parnami2022few, jung2023metric} have predominantly concentrated on UDKWS systems where an audio sample is used for pre-enrolling the keyword, known as query-by-example (QbyE) methods. The performance of QbyE methods is highly variable, mainly due to discrepancies between the pre-enrolled audio and the input spoken utterance. In response to the disparities in audio samples and to enhance user convenience, UDKWS systems have incorporated a method for text-based keyword enrollment. However, text lacks acoustic information compared to audio, making it challenging to reduce the distinctions between these two modalities~\cite{tan2021survey}.

To address this issue, current research in UDKWS with text-based enrollment predominantly focuses on reducing the discrepancy between audio and text modalities. Establishing a phoneme-to-vector database by converting phonemes into averages of frame-level audio embeddings effectively reduces the mismatch between audio and text embedding spaces on in-domain datasets~\cite{nishu2023flexible}. However, its performance in unseen domain datasets remains suboptimal. Other methods assess the similarity between audio and text embeddings using attention-based modules. Shin et al.~\cite{shin2022learning} leverage audio embeddings as the key and value, and text embeddings as the query to the cross-attention module to evaluate the similarity between two modalities at the utterance level. Lee et al.~\cite{lee2023phonmatchnet} suggest a self-attention-based framework that merges audio and text embeddings into a singular representation. 

To strengthen the correlation between two different modalities (audio and text), we present ParallelKWS, a UDKWS framework that adopts both self- and cross-attention mechanisms~\cite{vaswani2017attention}. The effectiveness of modality fusion using both self- and cross-attention has been reported in various deep learning fields, including speech emotion recognition~\cite{sun2021multimodal, yang2022contextual, luo2023cross} and feature matching~\cite{sarlin2020superglue, wang2022matchformer, lu2023paraformer}. However, it has not yet been explored in the context of keyword spotting. The self-attention module captures both inter- and intra-modal information by processing concatenated audio and text embeddings as input~\cite{lee2023phonmatchnet}. 
To enrich the inter-modal information influenced by each respective modality, ParallelKWS also incorporates two cross-attention modules using audio and text embeddings as their queries, respectively.

Furthermore, we propose a phoneme duration-based alignment loss as an auxiliary training objective to obtain a fine-grained alignment between the embeddings from audio and text modalities. We employ a pre-trained speech embedder as a component of the audio encoder. As this embedder is trained with phoneme-level connectionist temporal classification (CTC) loss~\cite{graves2006connectionist}, the model inherently provides phoneme duration information of the audio samples at frame-level. A target matrix, generated from this duration information, is utilised to enforce sequential correspondence between audio and text. By aligning the phonetic timing of the spoken words with the corresponding textual representation, this approach improves how the model associates varied speech patterns with their textual counterparts. Experimental results show that our approach outperforms comparable previous works on most benchmark datasets, and demonstrate the effectiveness of our proposed framework.

\begin{figure*}[ht]
\label{main_fig}
  \centering
  \includegraphics[width=0.95\linewidth]{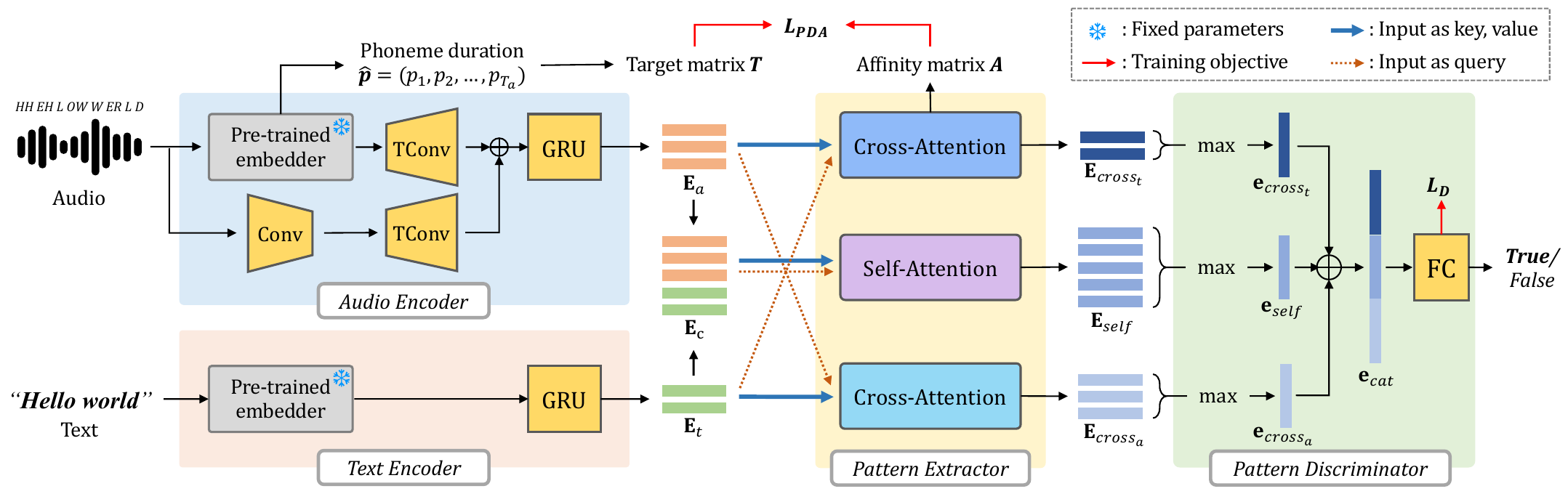}
  \caption{Overall framework of ParallelKWS. ``TConv" denotes ``Transposed convolution". ``max" and ``FC" stand for ``Max-pooling" and ``Fully-connected layer," respectively. Audio embedding $\mathbf{E}_a$ is used as query in one cross-attention module, with text embedding $\mathbf{E}_t$ serving as key and value. In the other cross-attention module, this setup is reversed. Concatenated embedding $\mathbf{E}_c$ is input to self-attention module. Outputs from each attention module are max-pooled and then concatenated. Finally, the concatenated output is passed through FC to produce the final logit.}
  \label{fig:overall_framework}
  \vspace{-4mm}
\end{figure*}

\section{Proposed Method}

In this section, we describe our proposed framework including model architecture and training objective. The overall framework is illustrated in~\Fref{fig:overall_framework}. It comprises two distinct encoders that capture features from the audio and text modalities. The framework also includes a pattern extractor, which combines these audio and text features, and a pattern discriminator responsible for determining the presence of the keyword.

\begin{table*}[!t]
\centering
\caption{Comparison of model performances and ablation study on the training objective. \\ 
The results for $\dag$ are as reported in prior works. The best results are in bold.
}
\resizebox{1.0\textwidth}{!}{\begin{tabular}{lcccccccc}
\toprule
\multirow{2}{*}{Method} & \multicolumn{4}{c|}{EER (\%)~$\downarrow$}                                      & \multicolumn{4}{c}{AUC (\%)~$\uparrow$}                                       \\ \cmidrule{2-9} 
                        & \multicolumn{1}{c}{G} & \multicolumn{1}{c}{Q} & \multicolumn{1}{c}{LP$_{\text{E}}$} & \multicolumn{1}{c|}{LP$_{\text{H}}$} & \multicolumn{1}{c}{G} & \multicolumn{1}{c}{Q} & \multicolumn{1}{c}{LP$_{\text{E}}$} & LP$_{\text{H}}$ \\
\midrule
\quad \textbf{\textit{Baselines}}              &   &    &  &   &    &   &   &   \\
CMCD~\cite{shin2022learning}$^{\dag}$            & \multicolumn{1}{c}{27.25}   & \multicolumn{1}{c}{12.15}   & \multicolumn{1}{c}{8.42}  & \multicolumn{1}{c|}{32.90}  & \multicolumn{1}{c}{81.06}   & \multicolumn{1}{c}{94.51}   & \multicolumn{1}{c}{96.70}  & 73.58  \\
FlexiKWS~\cite{nishu2023flexible}$^{\dag}$   & \multicolumn{1}{c}{14.05}   & \multicolumn{1}{c}{-}   & \multicolumn{1}{c}{0.8}  & \multicolumn{1}{c|}{18.4}  & \multicolumn{1}{c}{93.16}   & \multicolumn{1}{c}{-}   & \multicolumn{1}{c}{99.94}  & 89.2  \\
PhonMatchNet (re-impl.)~\cite{lee2023phonmatchnet}         & \multicolumn{1}{c}{14.04 {\scriptsize $\pm 1.08$}}   & \multicolumn{1}{c}{11.72 {\scriptsize $\pm 1.28$}}   & \multicolumn{1}{c}{0.48 {\scriptsize $\pm 0.03$}}  & \multicolumn{1}{c|}{18.77 {\scriptsize $\pm 0.22$}}  & \multicolumn{1}{c}{93.81 {\scriptsize $\pm 0.83$}}   & \multicolumn{1}{c}{95.50 {\scriptsize $\pm 0.84$}}   & \multicolumn{1}{c}{99.80 {\scriptsize $\pm 0.01$}}  & 88.01 {\scriptsize $\pm 0.16$}  \\
\midrule
\quad \textbf{\textit{Baselines (with additional data)}}              &   &    &  &   &    &   &   &   \\
FlexiKWS w/ neg.~\cite{nishu2023flexible}$^{\dag}$   & \multicolumn{1}{c}{13.45}   & \multicolumn{1}{c}{-}   & \multicolumn{1}{c}{1.7}  & \multicolumn{1}{c|}{14.4}  & \multicolumn{1}{c}{93.94}   & \multicolumn{1}{c}{-}   & \multicolumn{1}{c}{99.84}  & \bf{92.7}  \\
PhonMatchNet~\cite{lee2023phonmatchnet}$^{\dag}$   & \multicolumn{1}{c}{\bf{6.77}}   & \multicolumn{1}{c}{4.75}   & \multicolumn{1}{c}{2.80}  & \multicolumn{1}{c|}{18.82}  & \multicolumn{1}{c}{\bf{98.11}}   & \multicolumn{1}{c}{98.90}   & \multicolumn{1}{c}{99.29}  & 88.52  \\
\midrule
\quad \textbf{\textit{ParallelKWS (ours)}}              &   &    &  &   &    &   &   &   \\
Only detection loss                                                 & \multicolumn{1}{c}{8.78 {\scriptsize $\pm 0.27$}}   & \multicolumn{1}{c}{2.90 {\scriptsize $\pm 0.11$}}   & \multicolumn{1}{c}{0.11 {\scriptsize $\pm 0.01$}}  & \multicolumn{1}{c|}{14.80 {\scriptsize $\pm 0.10$}}  & \multicolumn{1}{c}{97.31 {\scriptsize $\pm 0.18$}}   & \multicolumn{1}{c}{99.66 {\scriptsize $\pm 0.03$}}   & \multicolumn{1}{c}{99.96 {\scriptsize $\pm 0.00$}}  & 91.29 {\scriptsize $\pm 0.10$}  \\
Detection loss + MM loss            & \multicolumn{1}{c}{8.02 {\scriptsize $\pm 0.25$}}   & \multicolumn{1}{c}{3.38 {\scriptsize $\pm 0.83$}}   & \multicolumn{1}{c}{0.13 {\scriptsize $\pm 0.01$}}  & \multicolumn{1}{c|}{15.46 {\scriptsize $\pm 0.13$}}  & \multicolumn{1}{c}{97.72 {\scriptsize $\pm 0.16$}}   & \multicolumn{1}{c}{99.50 {\scriptsize $\pm 0.24$}}   & \multicolumn{1}{c}{99.95 {\scriptsize $\pm 0.01$}}  & 90.72 {\scriptsize $\pm 0.15$}  \\
Detection loss + PDA loss                      & \multicolumn{1}{c}{7.78 {\scriptsize $\pm 0.40$}}   & \multicolumn{1}{c}{\bf{2.61} {\scriptsize $\pm 0.18$}}   & \multicolumn{1}{c}{\bf{0.09} {\scriptsize $\pm 0.01$}}  & \multicolumn{1}{c|}{\bf{14.36} {\scriptsize $\pm 0.07$}}  & \multicolumn{1}{c}{97.75 {\scriptsize $\pm 0.13$}}   & \multicolumn{1}{c}{{\bf 99.67 {\scriptsize $\pm 0.05$}}}   & \multicolumn{1}{c}{{\bf 99.97 {\scriptsize $\pm 0.01$}}}  & 91.68 {\scriptsize $\pm 0.10$}  \\
\bottomrule
\end{tabular}}
\label{main_tab}
\end{table*}
\subsection{Model Architecture}

\newpara{Audio encoder.}
The audio encoder is composed of a dual-path feature extractor~\cite{lee2023phonmatchnet} followed by a single GRU~\cite{cho2014learning} layer. 
One path of the feature extractor includes a pre-trained speech embedder, a small conformer~\cite{gulati2020conformer} optimised using phoneme-level CTC loss, and a 1-D transposed convolution with a kernel size of 5 and a stride of 4. Following \cite{nishu2023flexible}, the conformer is structured with 6 encoder layers, an encoder dimension of 144, a convolution kernel size of 3, and 4 attention heads.
The other path consists of a 1-D convolution with a kernel size of 3 and stride of 2, followed by a 1-D transposed convolution with a kernel size of 3 and stride of 2.
The outputs from both paths are concatenated along the feature dimension and then fed into a GRU layer, yielding the final audio embeddings. Audio embeddings are denoted as \(\mathbf{E}_a \in \mathbb{R}^{T_a \times d}\), where \(T_a\) and \(d\) represent the lengths of the audio features and the dimension of the embeddings, respectively. \(d\) is set to 128 in this study.

\newpara{Text encoder.}
To reduce the mismatch with the output of the phoneme-based audio encoder, we use a pre-trained grapheme-to-phoneme (G2P)~\cite{g2pE2019} model as a text encoder, followed by a single GRU layer.
The G2P embeddings are derived from the last hidden states of the encoder~\cite{lee2023phonmatchnet}.
The text embeddings are denoted as \( \mathbf{E}_t \in \mathbb{R}^{T_t \times d} \), where \(T_t\) refers to the lengths of the text features.

\newpara{Pattern extractor.}
To effectively fuse audio and text information, we construct a pattern extractor using the Parallel-attention, which combines cross-attention and self-attention modules in parallel.
The attention mechanism calculates the weighted sum of the values (\( V \)), based on the similarity scores between the queries (\( Q \)) and keys (\( K \)):
\begin{equation}
\text{Attn}(Q, K, V) = \text{Softmax}\left(\frac{QK^T}{\sqrt{d_k}}\right)V.
\end{equation}

\(\mathbf{E}_{cross_t}\) is the output embedding of the cross-attention module, with the text embedding \(\mathbf{E}_t\) as the query and the audio embedding \(\mathbf{E}_a\) as the key and value. Conversely, \(\mathbf{E}_{cross_a}\) is obtained from the other cross-attention module where the audio embedding is utilised as the query:

\begin{equation}
\mathbf{E}_{cross_t} = \text{Attn}(\mathbf{E}_t\mathbf{W}_t^Q, \mathbf{E}_a\mathbf{W}_a^K, \mathbf{E}_a\mathbf{W}_a^V),
\end{equation}
\begin{equation}
\mathbf{E}_{cross_a} = \text{Attn}(\mathbf{E}_a\mathbf{W}_a^Q, \mathbf{E}_t\mathbf{W}_t^K, \mathbf{E}_t\mathbf{W}_t^V).
\end{equation}

In the self-attention mechanism, the unimodal embeddings \( \mathbf{E}_a \) and \( \mathbf{E}_t \) are concatenated across the time dimension to form the concatenated embedding \( \mathbf{E}_c \), which is utilized as the query, key, and value to obtain the self-attention output \(\mathbf{E}_{self}\):

\begin{equation}
\mathbf{E}_{self} = \text{Attn}(\mathbf{E}_c\mathbf{W}_c^Q, \mathbf{E}_c\mathbf{W}_c^K, \mathbf{E}_c\mathbf{W}_c^V),
\end{equation}
where $\mathbf{W}^Q$, $\mathbf{W}^K$, and $\mathbf{W}^V$
are projection matrices of the query, key, and value, respectively. 

\newpara{Pattern discriminator.}
The pattern discriminator determines whether the keyword is detected. We first apply a max-pooling layer along the time axis to condense the outputs from both the cross- and self-attention modules. These condensed outputs are then concatenated along the feature axis to create the integrated features. Finally, we employ a fully-connected layer with a sigmoid activation. The process is summarised as follows:
\begin{equation}
\mathbf{e}_{cat} = \text{Concat}(\mathbf{e}_{cross_t}, \mathbf{e}_{cross_a}, \mathbf{e}_{self})
\end{equation}
\begin{equation}
\hat{y} = \sigma(\mathbf{W} \cdot \mathbf{e}_{cat} + \mathbf{b})
\end{equation}
where \(\mathbf{e}_{cross_t}\), \(\mathbf{e}_{cross_a}\), and \(\mathbf{e}_{self}\) in \(\mathbb{R}^d\) represent the condensed features of \(\mathbf{E}_{cross_t}\), \(\mathbf{E}_{cross_a}\), and \(\mathbf{E}_{self}\), and \(\mathbf{W}\), \(\mathbf{b}\), \(\sigma\) are the weights, biases, and the sigmoid function of the fully-connected layer, respectively.

\subsection{Training Objective}

\newpara{Phoneme duration-based alignment loss.}
We propose a novel training objective that enforces the model to learn the sequential correspondence between audio and text modalities based on phoneme duration information extracted from a pre-trained speech embedder. Inspired by~\cite{shin2022learning}, we align the sequential information from audio and text embeddings by matching the affinity matrix with the phoneme duration-based target matrix. Here, the attention map from the cross-attention module with the query of text embeddings is used as the affinity matrix. For negative pairs, we utilise a target matrix derived from normally distributed random noise, following~\cite{shin2022learning}.

For a given positive audio-text pair, $\hat{\boldsymbol{p}}=(p_1, p_2, ..., p_{T_a})$ represents a vector of phoneme predictions from the pre-trained speech embedder. As these predictions are at the frame level, the number of consecutive identical phoneme predictions likely contains information about the phoneme duration of the audio sample. We assign a group index to each $p_{i}$ in $\hat{\boldsymbol{p}}$, incrementing the index whenever a new phoneme prediction appears, thus grouping consecutive identical phoneme predictions. The resulting consecutive index vector can be denoted as 
$\hat{\boldsymbol{c}}=(c_1, c_2, ..., c_{T_a})$, where $c_{i}$ is defined as follows:
\begin{equation}
    c_i=
    \begin{cases}
        1, & \text{if}\ i=1 \\
        c_{i-1}, & \text{if}\ p_{i}=p_{i-1} \\
        c_{i-1}+1, & \text{if}\ p_{i}\neq p_{i-1}.
    \end{cases}
\end{equation}

Using the above index vector, we define matrix $\boldsymbol{D}=\left[ d_{ij} \right] \in \mathbb{R}^{T_{a} \times T_{t}}$, where $d_{ij} = j - c_{i}$ for all $i, j \in \mathbb{N}$, $1 \leq i \leq T_a$, and $1 \leq j \leq T_t$. The phoneme duration-based target matrix $\boldsymbol{T}=\left[ t_{ij} \right] \in \mathbb{R}^{T_{a} \times T_{t}}$ is obtained through the following equation.
\begin{equation}
\label{eq_target}
    x_{ij}=
    -\frac{(d_{ij} / T_t)^2}{2g^2}.
\end{equation}
\begin{equation}
    t_{ij}=
    \frac{\exp{(x_{ij})}}{{\sum_{i}{\exp{(x_{ij})}}}}.
\end{equation}
Here, $g$ is a hyperparameter that determines the gradient of the exponential function and is set to $0.1$ in this work. The phoneme duration-based alignment loss is defined as the mean square error between the affinity matrix $\boldsymbol{A}$ and the phoneme duration-based target matrix $\boldsymbol{T}$:
\begin{equation}
    L_{PDA}={||\boldsymbol{A} - \boldsymbol{T}||}^2.
\end{equation}

\newpara{Detection loss.}
To assess if the input audio sample and the input text correspond to the same keyword, we use binary cross-entropy loss on the logits from the pattern discriminator. Since this detection loss addresses both the entire audio sample and the phonemes within it, the network is trained to recognise similarities between audio and text at the level of entire utterances.

\begin{equation}
L_{D} = -(y \cdot \log \hat{y} + (1 - y) \cdot \log(1 - \hat{y})),
\end{equation}
where $\hat{y}$ and $y$ denote the predicted probability and the ground truth label, respectively.

Finally, we formulate the overall loss ($L_{total}$) as follows:

\begin{equation}
L_{total}=\lambda \cdot L_{PDA}+L_{D},
\end{equation}
where $\lambda$ is a weight factor, and is set to 0.3.

\section{Experiments}

\subsection{Datasets and Evaluation Methods}
We employ the LibriPhrase~\cite{shin2022learning} dataset, which comprises phrases ranging from 1 to 4 words, and is divided into a training set and a test set, derived from distinct splits of the LibriSpeech~\cite{panayotov2015librispeech} dataset: \textit{train-clean} and \textit{train-other}. We use 800k phrases for training, evenly distributed with 200k phrases for each word length, in line with~\cite{nishu2023flexible, shin2022learning, nishu23_interspeech}. Additionally, we use LibriSpeech \textit{train-clean} dataset along with LibriPhrase training set to train the conformer with phoneme-level CTC loss. This training involves an initial phase on the LibriSpeech \textit{train-clean} dataset, followed by fine-tuning using shorter audio segments from the LibriPhrase training set. Input audio data augmentation is performed using various noises from the MUSAN~\cite{snyder2015musan} dataset and room impulse response filters. The entire network is then trained on the LibriPhrase training set, without updating the parameters in the pre-trained conformer and G2P model. Audio features are extracted using 80-channel filterbanks with a 25ms window and a 10ms frame shift.

The LibriPhrase test set is categorised based on the Levenshtein distance~\cite{levenshtein1966binary} between negative pairs, where a lower distance indicates higher phonetic similarity and greater difficulty in discrimination. The test set with hard negative pairs and easy negative pairs are labeled as LibriPhrase-hard (\textbf{LP$_{\textbf{H}}$}) and LibriPhrase-easy (\textbf{LP$_{\textbf{E}}$}), respectively. For evaluation, four distinct KWS benchmark datasets are used: LibriPhrase-easy, LibriPhrase-hard, Google Speech Commands V1 (\textbf{G})~\cite{warden2018speech}, and Qualcomm Keyword Speech (\textbf{Q})~\cite{kim2019query}, with the official split for \textbf{LP$_{\textbf{E}}$} and \textbf{LP$_{\textbf{H}}$} as provided in~\cite{shin2022learning}. For \textbf{G} and \textbf{Q}, we adhere to the testing protocol in~\cite{lee2023phonmatchnet} to maintain fairness in comparison, considering all keywords except the anchor keyword as negatives. We report the Equal Error Rate (EER) and Area Under the ROC Curve (AUC) scores for each benchmark dataset. We present the average performance and standard deviation across three experiments, each conducted with a distinct random seed for reliability.

\subsection{Implementation Details}

The network is optimised for 100 epochs using the Adam optimizer~\cite{kingma2014adam}, set to a fixed learning rate of $1\mathrm{e}$-${3}$. For evaluation, we select the model with the lowest EER on the test sets. We establish the batch size at 2048, and the training process takes approximately one day on a single A5000 GPU which has a memory size of 24GB. The framework for our model is implemented using the PyTorch library.

\begin{table}[t!]
\centering
\caption{Effectiveness of the attention modules in pattern extractor.
 The number of parameters (\# params.) only reflects the trainable parameters. Pre-trained embedders contain 2.33M parameters. The best results are in bold.
}
\resizebox{1.0\linewidth}{!}{\begin{tabular}{cc|c|cccc}
\toprule
\multicolumn{3}{c|}{Model}  & \multicolumn{4}{c}{EER (\%)~$\downarrow$}                                                    \\ \midrule
Self        & Cross         & \# params. & \multicolumn{1}{c}{G}   & \multicolumn{1}{c}{Q}   & \multicolumn{1}{c}{LP$_{\text{E}}$} & LP$_{\text{H}}$ \\ \midrule
\cmark       &             & $0.55M$ & \multicolumn{1}{c}{9.38 {\scriptsize $\pm 0.32$}}   & \multicolumn{1}{c}{4.81 {\scriptsize $\pm 1.57$}}   & \multicolumn{1}{c}{0.23 {\scriptsize $\pm 0.03$}}  & 18.56 {\scriptsize $\pm 0.30$}  \\
            & \cmark       & $0.61M$ & \multicolumn{1}{c}{9.32 {\scriptsize $\pm 0.21$}}   & \multicolumn{1}{c}{3.24 {\scriptsize $\pm 0.63$}}   & \multicolumn{1}{c}{0.16 {\scriptsize $\pm 0.02$}}  & 15.80 {\scriptsize $\pm 0.16$}  \\
\cmark       & \cmark       & $0.68M$ & \multicolumn{1}{c}{{\bf 8.78} {\scriptsize $\pm 0.27$}}   & \multicolumn{1}{c}{{\bf 2.90} {\scriptsize $\pm 0.11$}}   & \multicolumn{1}{c}{{\bf 0.11} {\scriptsize $\pm 0.01$}}  & {\bf 14.80} {\scriptsize $\pm 0.10$}  \\ 
\bottomrule
\end{tabular}}
\label{model_tab}
\end{table}
\section{Results}

\subsection{Comparison with Baselines}

In~\Tref{main_tab}, we report the performance of our proposed framework, ParallelKWS, alongside that of existing baselines. CMCD~\cite{shin2022learning} and FlexiKWS~\cite{nishu2023flexible} utilise the same training dataset as our study. However, PhonMatchNet~\cite{lee2023phonmatchnet} employs a KWS model pre-trained on various external domain data ($200M$ audio clips) collected from YouTube~\cite{lin2020training} as a speech embedder. To ensure a fair comparison, we also present the performance of PhonMatchNet re-implemented using the same speech embedder as ours.
When trained on the same dataset, ParallelKWS outperforms all existing baselines in terms of both EER and AUC scores. Notably, our method significantly improves performance on the \textbf{LP$_{\textbf{E}}$} dataset by 99.0\%, 89.2\%, and 82.0\% over CMCD, FlexiKWS, and the re-implemented PhonMatchNet, respectively.

We also compare our model with those trained using additional datasets. FlexiKWS uses phonetically confusable keywords as additional training data, labeled as \textit{FlexiKWS w/ neg.} in ~\Tref{main_tab}. Nevertheless, ParallelKWS demonstrates improved performance on all test sets, except for a slight (1.1\%) decrease in AUC score on the \textbf{LP$_{\textbf{H}}$} dataset. Compared to PhonMatchNet, which includes a speech embedder pre-trained on large-scale external data and employs phoneme-level detection loss to enhance the capability of distinguishing similar pronunciations, ParallelKWS shows improved performance on the datasets except for \textbf{G}. Especially on the \textbf{LP$_{\textbf{H}}$} dataset, our framework demonstrates a significant improvement of 23.7\%. Through the comparison of performance with the baselines, we highlight the effectiveness of ParallelKWS in capturing data dependencies from unseen domains (\textbf{G} and \textbf{Q}) without additional training processes, while maintaining its capability with in-domain data (\textbf{LP$_{\textbf{E}}$} and \textbf{LP$_{\textbf{H}}$}).

\subsection{Ablation Study}

\newpara{Effectiveness of parallel-attention architecture.}
We demonstrate the impact of parallel-attention architecture through ablation studies on the attention modules within the pattern extractor.
As shown in \Tref{model_tab}, using either self-attention or cross-attention results in performance degradations across all test sets compared to their parallel connection. Notably, parallel-attention significantly improves performance in the \textbf{LP$_{\textbf{H}}$} dataset, which contains marginally distinguishable pronunciations, as well as in the test sets from unseen domains, \textbf{G} and \textbf{Q}. These results indicate that the parallel-attention architecture precisely captures both inter- and intra-modal information from audio and text, effectively merging the two modalities. 

\newpara{Effectiveness of phoneme duration-based alignment loss.}
We conduct an ablation study to assess the functionality of the proposed phoneme duration-based alignment loss (PDA loss). This study aims to confirm the effectiveness of incorporating phoneme duration information. We compare our approach with the monotonic matching loss (MM loss) proposed in~\cite{shin2022learning}, applying it to our architecture. The monotonic matching approach aligns the affinity matrix with a target matrix that is organised in a monotonic order across the audio and text sequences. However, this target matrix lacks intrinsic duration-related information.

As indicated in the seventh row of \Tref{main_tab}, using MM loss leads to decreased performance in the \textbf{Q}, \textbf{LP$_{\textbf{E}}$}, and \textbf{LP$_{\textbf{H}}$} datasets, while it shows improvement in the \textbf{G} dataset. In contrast, using PDA loss results in an EER reduction of 11.4\%, 10.0\%, 21.2\%, and 3.0\% in the \textbf{G}, \textbf{Q}, \textbf{LP$_{\textbf{E}}$}, and \textbf{LP$_{\textbf{H}}$} datasets, respectively. This result emphasises that aligning audio and text sequences in the affinity matrix based on phoneme duration encourages the preceding modules to be trained with an enhanced capability to capture duration-related sequential information.

\section{Conclusion}
In this paper, we introduce a framework for user-defined keyword spotting with text-based enrollment, effectively integrating audio and text information. Our framework employs cross- and self-attention modules in a parallel architecture to capture both inter- and intra-modal information, thus improving the capability of the model to fuse audio and text modalities. We also implement an alignment loss that utilises phoneme duration information, derived from the pre-trained speech embedder, to enhance the alignment of sequential information between audio and text embeddings. Experimental results show that our framework outperforms previous models on most test sets, achieving this without any training on data from external domains.


\bibliographystyle{IEEEtran}
\bibliography{shortstrings,refs}

\begin{thebibliography}{10}
\providecommand{\url}[1]{#1}
\csname url@samestyle\endcsname
\providecommand{\newblock}{\relax}
\providecommand{\bibinfo}[2]{#2}
\providecommand{\BIBentrySTDinterwordspacing}{\spaceskip=0pt\relax}
\providecommand{\BIBentryALTinterwordstretchfactor}{4}
\providecommand{\BIBentryALTinterwordspacing}{\spaceskip=\fontdimen2\font plus
\BIBentryALTinterwordstretchfactor\fontdimen3\font minus \fontdimen4\font\relax}
\providecommand{\BIBforeignlanguage}[2]{{%
\expandafter\ifx\csname l@#1\endcsname\relax
\typeout{** WARNING: IEEEtran.bst: No hyphenation pattern has been}%
\typeout{** loaded for the language `#1'. Using the pattern for}%
\typeout{** the default language instead.}%
\else
\language=\csname l@#1\endcsname
\fi
#2}}
\providecommand{\BIBdecl}{\relax}
\BIBdecl

\bibitem{chen2014small}
G.~Chen, C.~Parada, and G.~Heigold, ``Small-footprint keyword spotting using deep neural networks,'' in \emph{Proc. ICASSP}, 2014.

\bibitem{hou2019region}
J.~Hou, Y.~Shi, M.~Ostendorf, M.-Y. Hwang, and L.~Xie, ``Region proposal network based small-footprint keyword spotting,'' \emph{IEEE Signal Processing Letters}, 2019.

\bibitem{tang2018deep}
R.~Tang and J.~Lin, ``Deep residual learning for small-footprint keyword spotting,'' in \emph{Proc. ICASSP}, 2018.

\bibitem{berg2021keyword}
A.~Berg, M.~O'Connor, and M.~T. Cruz, ``Keyword transformer: A self-attention model for keyword spotting,'' in \emph{Proc. Interspeech}, 2021.

\bibitem{huh2021metric}
J.~Huh, M.~Lee, H.~Heo, S.~Mun, and J.~S. Chung, ``Metric learning for keyword spotting,'' in \emph{IEEE Spoken Language Technology workshop}, 2021.

\bibitem{chen2015query}
G.~Chen, C.~Parada, and T.~N. Sainath, ``Query-by-example keyword spotting using long short-term memory networks,'' in \emph{Proc. ICASSP}, 2015.

\bibitem{huang2021query}
J.~Huang, W.~Gharbieh, H.~S. Shim, and E.~Kim, ``Query-by-example keyword spotting system using multi-head attention and soft-triple loss,'' in \emph{Proc. ICASSP}, 2021.

\bibitem{mazumder21_interspeech}
M.~Mazumder, C.~Banbury, J.~Meyer, P.~Warden, and V.~J. Reddi, ``{Few-Shot Keyword Spotting in Any Language},'' in \emph{Proc. Interspeech}, 2021.

\bibitem{parnami2022few}
A.~Parnami and M.~Lee, ``Few-shot keyword spotting with prototypical networks,'' in \emph{International Conference on Machine Learning Technologies}, 2022.

\bibitem{jung2023metric}
J.~Jung, Y.~Kim, J.~Park, Y.~Lim, B.-Y. Kim, Y.~Jang, and J.~S. Chung, ``Metric learning for user-defined keyword spotting,'' in \emph{Proc. ICASSP}, 2023.

\bibitem{tan2021survey}
X.~Tan, T.~Qin, F.~Soong, and T.-Y. Liu, ``A survey on neural speech synthesis,'' \emph{arXiv preprint arXiv:2106.15561}, 2021.

\bibitem{nishu2023flexible}
K.~Nishu, M.~Cho, P.~Dixon, and D.~Naik, ``Flexible keyword spotting based on homogeneous audio-text embedding,'' in \emph{Proc. ICASSP}, 2023.

\bibitem{shin2022learning}
H.-K. Shin, H.~Han, D.~Kim, S.-W. Chung, and H.-G. Kang, ``Learning audio-text agreement for open-vocabulary keyword spotting,'' in \emph{Proc. Interspeech}, 2022.

\bibitem{lee2023phonmatchnet}
Y.-H. Lee and N.~Cho, ``Phonmatchnet: phoneme-guided zero-shot keyword spotting for user-defined keywords,'' in \emph{Proc. Interspeech}, 2023.

\bibitem{vaswani2017attention}
A.~Vaswani, N.~Shazeer, N.~Parmar, J.~Uszkoreit, L.~Jones, A.~N. Gomez, {\L}.~Kaiser, and I.~Polosukhin, ``Attention is all you need,'' in \emph{Proc. NeurIPS}, 2017.

\bibitem{sun2021multimodal}
L.~Sun, B.~Liu, J.~Tao, and Z.~Lian, ``Multimodal cross-and self-attention network for speech emotion recognition,'' in \emph{Proc. ICASSP}, 2021.

\bibitem{yang2022contextual}
D.~Yang, S.~Huang, Y.~Liu, and L.~Zhang, ``Contextual and cross-modal interaction for multi-modal speech emotion recognition,'' \emph{IEEE Signal Processing Letters}, vol.~29, pp. 2093--2097, 2022.

\bibitem{luo2023cross}
M.~Luo, H.~Phan, and J.~Reiss, ``cross-modal fusion techniques for utterance-level emotion recognition from text and speech,'' in \emph{Proc. ICASSP}, 2023.

\bibitem{sarlin2020superglue}
P.-E. Sarlin, D.~DeTone, T.~Malisiewicz, and A.~Rabinovich, ``Superglue: Learning feature matching with graph neural networks,'' in \emph{Proc. CVPR}, 2020.

\bibitem{wang2022matchformer}
Q.~Wang, J.~Zhang, K.~Yang, K.~Peng, and R.~Stiefelhagen, ``Matchformer: Interleaving attention in transformers for feature matching,'' in \emph{Proc. ACCV}, 2022.

\bibitem{lu2023paraformer}
X.~Lu, Y.~Yan, B.~Kang, and S.~Du, ``Paraformer: Parallel attention transformer for efficient feature matching,'' in \emph{Proc. AAAI}, 2023.

\bibitem{graves2006connectionist}
A.~Graves, S.~Fern{\'a}ndez, F.~Gomez, and J.~Schmidhuber, ``Connectionist temporal classification: labelling unsegmented sequence data with recurrent neural networks,'' in \emph{Proc. ICML}, 2006.

\bibitem{cho2014learning}
K.~Cho, B.~Van~Merri{\"e}nboer, C.~Gulcehre, D.~Bahdanau, F.~Bougares, H.~Schwenk, and Y.~Bengio, ``Learning phrase representations using rnn encoder-decoder for statistical machine translation,'' in \emph{Proc. EMNLP}, 2014.

\bibitem{gulati2020conformer}
A.~Gulati, J.~Qin, C.-C. Chiu, N.~Parmar, Y.~Zhang, J.~Yu, W.~Han, S.~Wang, Z.~Zhang, Y.~Wu \emph{et~al.}, ``Conformer: Convolution-augmented transformer for speech recognition,'' in \emph{Proc. Interspeech}, 2020.

\bibitem{g2pE2019}
K.~Park and J.~Kim, ``g2pe,'' \url{https://github.com/Kyubyong/g2p}, 2019.

\bibitem{panayotov2015librispeech}
V.~Panayotov, G.~Chen, D.~Povey, and S.~Khudanpur, ``Librispeech: an asr corpus based on public domain audio books,'' in \emph{Proc. ICASSP}, 2015.

\bibitem{nishu23_interspeech}
K.~Nishu, M.~Cho, and D.~Naik, ``{Matching Latent Encoding for Audio-Text based Keyword Spotting},'' in \emph{Proc. Interspeech}, 2023.

\bibitem{snyder2015musan}
D.~Snyder, G.~Chen, and D.~Povey, ``Musan: A music, speech, and noise corpus,'' \emph{arXiv preprint arXiv:1510.08484}, 2015.

\bibitem{levenshtein1966binary}
V.~I. Levenshtein \emph{et~al.}, ``Binary codes capable of correcting deletions, insertions, and reversals,'' in \emph{Soviet physics doklady}, 1966.

\bibitem{warden2018speech}
P.~Warden, ``Speech commands: A dataset for limited-vocabulary speech recognition,'' \emph{arXiv preprint arXiv:1804.03209}, 2018.

\bibitem{kim2019query}
B.~Kim, M.~Lee, J.~Lee, Y.~Kim, and K.~Hwang, ``Query-by-example on-device keyword spotting,'' in \emph{IEEE Automatic Speech Recognition and Understanding Workshop}, 2019.

\bibitem{kingma2014adam}
D.~P. Kingma and J.~L. Ba, ``Adam: A method for stochastic optimization,'' in \emph{Proc. ICLR}, 2014.

\bibitem{lin2020training}
J.~Lin, K.~Kilgour, D.~Roblek, and M.~Sharifi, ``Training keyword spotters with limited and synthesized speech data,'' in \emph{Proc. ICASSP}, 2020.

\end{thebibliography}

\end{document}